\documentclass[journal]{IEEEtran}
\ifCLASSINFOpdf
\else
\fi

\usepackage[utf8]{inputenc}
\usepackage{amsmath}
\usepackage{mathtools}
\usepackage{braket}
\usepackage{graphicx}
\graphicspath{{./figures/}}
\usepackage{amsfonts}
\usepackage[ruled,vlined,linesnumbered]{algorithm2e}
\let\oldnl\nl
\newcommand{\nonl}{\renewcommand{\nl}{\let\nl\oldnl}}

\usepackage{csquotes}
\usepackage{hhline}
\usepackage{amssymb}
\usepackage{listings}
\usepackage{color}
\definecolor{codegreen}{rgb}{0,0.6,0}
\definecolor{codegray}{rgb}{0.5,0.5,0.5}
\definecolor{codepurple}{rgb}{0.58,0,0.82}
\definecolor{backcolour}{rgb}{0.95,0.95,0.92}
\usepackage{subcaption}
\usepackage{titlesec}

\usepackage{dcolumn}
\usepackage{tabularx}
\usepackage{hyperref}
\hypersetup{
    colorlinks=true,
    linktoc=all,
    linkcolor=black,
    citecolor=magenta,
}
\usepackage{longtable}
\usepackage{braket}
\usepackage{orcidlink}
\newcolumntype{C}{>{\centering\arraybackslash}X}
\SetAlFnt{\small}

\usepackage[font=small]{caption}
\begin{document}

%

\title{\huge \bf A Blockchain-based Quantum Binary Voting for\newline Decentralized IoT Towards Industry 5.0}
\author{Utkarsh~Azad\,\orcidlink{0000-0001-7020-0305}\thanks{U.~Azad was with Center for Computational Natural Sciences and Bioinformatics (CCNSB), International Institute of Information Technology Hyderabad, TS, India at the time of manuscript preparation and is currently with Xanadu Quantum Technologies Inc., Toronto, ON, Canada (e-mail: utkarsh.azad@research.iiit.ac.in).}, Bikash~K.~Behera\,\orcidlink{0000-0003-2629-3377}\thanks{B.~K. Behera is with Bikash's Quantum (OPC) Pvt. Ltd., Mohanpur, WB, 741246 India, and Department of Physical Sciences, Institute of Science Education and Research Kolkata, Mohanpur, WB, India.}, Houbing Song\,\orcidlink{0000-0003-2631-9223}\thanks{Houbing Song is with Department of Information Systems, University of Maryland, Baltimore County, Baltimore, MD, USA.}, and Ahmed~Farouk\,\orcidlink{0000-0001-8702-7342}\thanks{A.~Farouk is with Department of Computer Science, Faculty of Computers and Artificial Intelligence, South Valley University, Hurghada, Egypt.}}%
\maketitle
\thispagestyle{plain}
\pagestyle{plain}

\begin{abstract}
Industry 5.0 depends on intelligence, automation, and hyperconnectivity operations for effective and sustainable human-machine collaboration. Pivotal technologies like the Internet of Things (IoT) enable this by facilitating connectivity and data-driven decision-making between cyber-physical devices. As IoT devices are prone to cyberattacks, they can use blockchain to improve transparency in the network and prevent data tampering. However, in some cases, even blockchain networks are vulnerable to Sybil and 51\% attacks. This has motivated the development of quantum blockchains that are more resilient to such attacks as they leverage post-quantum cryptographic protocols and secure quantum communication channels. In this work, we develop a quantum binary voting algorithm for the IoT-quantum blockchain frameworks that enables inter-connected devices to reach a consensus on the validity of transactions, even in the presence of potential faults or malicious actors. The correctness of the voting protocol is provided in detail, and the results show that it guarantees the achievement of a consensus securely against all kinds of significant external and internal attacks concerning quantum bit commitment, quantum blockchain, and quantum Byzantine agreement. We also provide an implementation of the voting algorithm with the quantum circuits simulated on the IBM Quantum platform and Simulaqron library.
\end{abstract}

\begin{IEEEkeywords}
Industry 5.0, Decentralized IoT, Binary voting, Bit Commitment, Quantum Blockchain
\end{IEEEkeywords}
\section{INTRODUCTION}

The Internet of Things (IoT) refers to the interaction between the physical and digital worlds through a network of interconnected objects such as smart devices, actuators, sensors, radio-frequency identification (RFID) tags, etc. Within the novel paradigm of IoT, these objects interact with each other via unique addressing schemes to share data and equip themselves with ubiquitous intelligence, which helps them to carry out complex tasks without human intervention. Initially introduced as part of Industry $4.0$ to distribute the workload over decentralized networks of devices, it has also proven to play a crucial role in fostering Industry $5.0$, especially in manufacturing, transportation, and healthcare \cite{5gBC}.

Typical decentralized approaches for implementing IoT networks had problems regarding reliability, scalability, and traceability of the data workflow on distributed nodes in the network \cite{bib_Hardwick}. One possible solution for these problems is to use the blockchain that records transactions securely and transparently on a distributed network. However, ensuring all IoT devices in the blockchain network agree on the validity of transactions and can coordinate effectively is tricky due to the resource constraints of IoT devices that inhibit employing heavyweight consensus mechanisms based on Proof of Work \cite{Sriman2020}. Moreover, the security of data integrity and communications on traditional blockchain networks is dependent on classical cryptography tasks that are threatened by the development of fault-tolerant quantum computers that are speculated to be able to perform critical tasks like factoring efficiently in the future \cite{Gidney2021, Bova2021}. A feasible approach to tackle the latter is to use the quantum blockchain, which would use quantum mechanical principles such as entanglement along with quantum secure communication and cryptography protocols to enhance the security and reliability of the blockchain in these respects \cite{Wang2022, S2024}.

In this work, we demonstrate a possible solution for the former, i.e., a novel self-tallying quantum binary voting algorithm with internal auditing for assisting in leader election and reaching a consensus on quantum blockchain networks. The key contributions of this paper are threefold. First, we provide a complete algorithmic description of our self-tallying quantum consensus protocols for decentralized IoT environments assuming that each pair of nodes (devices) is connected via an authenticated quantum channel along with a classical channel on a quantum blockchain, where non-participating nodes can independently audit the entire voting process and hence improving the work done in \cite{bib_Sun}. Second, the security analysis of the proposed has been done to prove its resilience against various attacks, both internal and external, and satisfies the requirements such as anonymity, binding, non-reusability, verifiability, self-tallying, etc. Finally, the consensus subroutine of the protocol has been implemented on the IBM Quantum platform \cite{bib_IBM} by designing the required quantum circuits and its ballot commitment subroutine has been verified via the Simulaqron library \cite{bib_simqron}.

\section{Related Work}\label{QVP:Sec2}

This work targets various key areas that have remained disjointed until recently. In this regard, we encourage users to first take a look at \cite{Mocnej2018} for decentralized IoT network, \cite{Tripathi2023} for blockchain technologies, \cite{9781107002173} for quantum computing and \cite{Kho2022} for electronic voting.

\subsection{Blockchain in IoT}\label{BIoT}
An IoT ecosystem is a system of systems with many technological and physical components interacting via the internet. Each component generates, shares, and analyzes data to perform a common goal interconnectedly. Due to its secure, decentralized, and autonomous capabilities, Blockchain technology can support IoT ecosystems in data decentralization, transparency, verifiability, consensus, and security \cite{bib_Cono, bib_Novo}. However, such integration has challenges, including scalability, security, interoperability, and regulation of both technologies as shown in \cite{bib_Reyna}. We refer readers to \cite{bib_Kouicem}, which has studied the IoT systems with their security challenges in further detail and presents blockchain-based solutions for them.

\subsection{Blockchain-based Voting Protocols} \label{BbVP}
Various blockchain-based voting protocols that employ classical techniques to reduce the involvement of the third party and achieve transparency have been proposed. For example, an approach has been proposed using non-broadcasting blocks where two blocks might be constructed but one remains non-broadcasted. The non-broadcasted block can be disseminated whenever desired, ensuring the nominee's choices are secure until the results are computed \cite{bib_Andrew}. Another key concern is ensuring voters' privacy and anonymity, which can be addressed by utilizing a central authority that grants voter eligibility \cite{bib_Liu}. At the same time, the voting protocol allows multiple attempts to vote with replacement to remove coercion \cite{bib_Hardwick}. Similarly, public blockchains that are visible to everyone and have no centralized authority might be used \cite{bib_V1, bib_V2}.

\subsection{Quantum-based Voting Protocols} \label{VRP}
Unconditionally secure voting protocols can be built by using quantum cryptography protocols. However, they should satisfy anonymity, receipt-free, binding, non-reusability, verifiability, eligibility, self-tally, and fairness requirements to be reliable and useful.
Several quantum protocols for electronic voting have recently been developed. For example, a voting system based on controlled teleportation has been proposed \cite{bib_prev4}, which meets all of the above characteristics but requires more simplicity to implement it. Here, we implement an enhanced voting protocol that satisfies all the aforementioned critical requirements while also being resilient to internal and external attacks. 

\begin{figure*}[!ht]
    \centering
    \includegraphics[width=\linewidth]{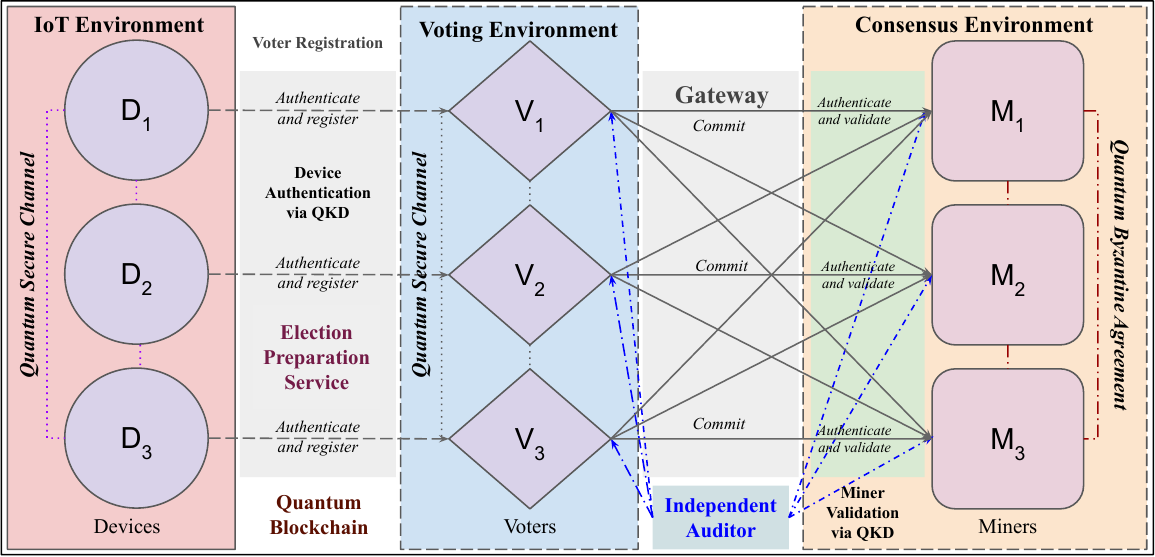}
    \caption{The system model of our blockchain-based quantum binary voting in a decentralized IoT. In our context, the protocol will have six components: (i) an IoT environment, which constitutes all the interconnected devices, (ii) an election preparation service (EPS), which authenticates and verifies devices as voters and maintains a voter list, (iii) voting environment, which constitutes all those devices that are authenticated and registered as voters (${V}$) by EPS, (iv) consensus environment, which constitutes of miners (${M}$) who take the masked ballots and reach a consensus on the result (v) the quantum-secure communication gateway which connects the voting and consensus environment, and (vi) an independent auditor, which can scrutinize the voting process for any complacency. The voters communicate with each other to prepare their masked ballots and then commit them to the miners residing on the same or different quantum blockchain. Then, miners authenticate the voters and their masked ballots via QKD protocol. Finally, on successful verification, they reach a consensus on the masked ballots via the quantum Byzantine agreement (QBA) protocol.}
    \label{fig:fig1}
\end{figure*}

\section{Preliminaries}\label{QVP:Sec3}

\subsection{Cheat Sensitive Quantum Bit Commitment}
In a fair decision-making process, each participant must be allowed to commit to a chosen value or statement while keeping it hidden from others, which they can reveal later for tallying and verification. The bit commitment scheme can legislate such a process while tagging any deviation from the fair process as a cheating attempt. If the probability of detecting these cheating attempts is non-zero, then it refers to such a commitment scheme as a cheat-sensitive quantum bit commitment (CSQBC) protocol \cite{bib_commit, Sun2020}.

In the context of the voting protocol in a blockchain environment, the nodes generally can act as voters $\in {V}$ and miners $\in {M}$. The voter V$_i$ uses CSQBC for committing $2k$ bits of information to miner M$_i$ which provides better security against attacks when compared to other bit commitment schemes. The whole protocol can be broken down into three phases, each of which would require the following two common parameters:

\begin{itemize}
\itemsep0em
\item \textit{m} $\Rightarrow$ The number of balanced-uniform sequences prepared by each M$_{j}$ to prove its integrity to V$_{i}$.
\item \textit{n}\ \ \,$\Rightarrow$ The length of each such sequence which is sent to each voter V$_{i}$ from each miner M$_{j}$.
\end{itemize} 

By choosing the values of these two parameters appropriately, the probability of failures involved can be reduced to be as small as required, i.e., causing the protocol to be probabilistically secure while making no assumptions about the intelligence or the computing power of miners or voters. 

\subsection{Quantum Byzantine Agreement}
In a distributed system, it is essential to reach reliability which generally requires all fault-free components of the system to agree on a common value, even if some components are corrupted. The problem of how to use such an agreement is known as the Byzantine agreement (BA) problem \cite{bib_qba}. 
We implement a quantum solution to the slightly weaker Byzantine agreement problem,where if one of the generals is complicit, the rest of the honest generals may or may not reach a consensus. This can be achieved by using entanglement distribution either to reach an agreement or disagreement by distributing many qutrit triplets $\ket{\psi_{j}}$ amongst the parties.

\subsection{Quantum Secure Communication protocol}
A communication protocol allows transmitting information via any variation of a physical quantity between two parties, and its security mainly depends upon the used key, which can be broken. Therefore, the quantum key distribution (QKD) protocol allows the unconditional secure transmission of random binary keys between the parties \cite{bib_qkd}. The security of these keys will depend upon forging the information encoded in non-orthogonal states. Both voters and miners can communicate via such a protocol using quantum secure communication (QSC). The required authentication steps for the nodes on the blockchain network are achieved via the QKD protocol on these channels, as shown in Fig. \ref{fig:fig1}. These include authenticating devices as voters by the election preparation service and the miners’ validation before the opening phase.

\subsection{Swapping}
The main challenge of protocol implementation on IBMQ systems is constrained by the number of qubits and their connectivity. As a result, we postulate the presence of a quantum secure communication channel through which voters and miners can interact. In our implementations, we rely on using swap gates as an analogous quantum circuit operation to implement the whole procedure for transferring information from voter V$_{i}$ to miner M$_{j}$.

\section{The Quantum Binary Voting System Model}\label{QVP:Sec4}

\subsection{System Model}

The proposed framework of the quantum binary voting system model in a decentralized IoT setting, which is blockchain-based and self-tallying, is shown in Fig. \ref{fig:fig1}. The system consists of six components: (i) devices in the IoT environment, (ii) an election preparation service (EPS), (iii) voters in the voting environment, (iv) miners in the consensus environment, (v) the quantum-secure communication gateway connects the voting and consensus environment, and (vi) an independent auditor.  The blockchain nodes can be smart devices, actuators, sensors, or any other IoT devices interconnected within the system. To qualify as a voter, each device must register and authenticate itself via EPS onto the network before releasing its votes through the gateway to the blockchain. Similarly, EPS will elect miners from the nodes on the same quantum blockchain that are not participating as voters. Once authentication and registration are completed, nodes that qualify as voters can communicate with other devices to prepare their masked ballots. After their preparation, ballots are sent to the mining nodes, which can authenticate them and determine the result by reaching a consensus. Note that the quantum blockchain leveraged in the model should be either a private or consortium blockchain (according to different voting scenarios) rather than a public blockchain so that participation could be further \textit{permissioned}. Furthermore, the whole system’s security depends on the quantum blockchain and quantum-secure communication gateway and the nodes that are neither participating as voters nor as miners can act as independent auditors to scrutinize the voting process for any complacency as they can interact with voters and miners.

\subsection{System Operations}\label{QVP:Sec4:SO}
To describe the operations of the binary voting protocol, we concentrate only on the voting and consensus environments, where there exists a network of voters (${V}$) and miners (${M}$), respectively. It is assumed that there exist quantum secure channels for their communication, and all voters on the blockchain use them to distribute a voting matrix $\mathbf{V}$ whose preparation is discussed below. Using this matrix, every voter $\in {V}$ generates a masked ballot which is then committed to all the miners in the same or different blockchain via a commitment protocol. Finally, all miners $\in {M}$ use the Byzantine agreement protocol to reach a consensus on the masked ballots to obtain the results eventually. The entire operation scheme is represented in Fig. \ref{fig:fig1} with each of the described components. We shall now discuss the main steps of our voting protocol.

\begin{figure}[t]
\centering
\includegraphics[width=0.98\linewidth]{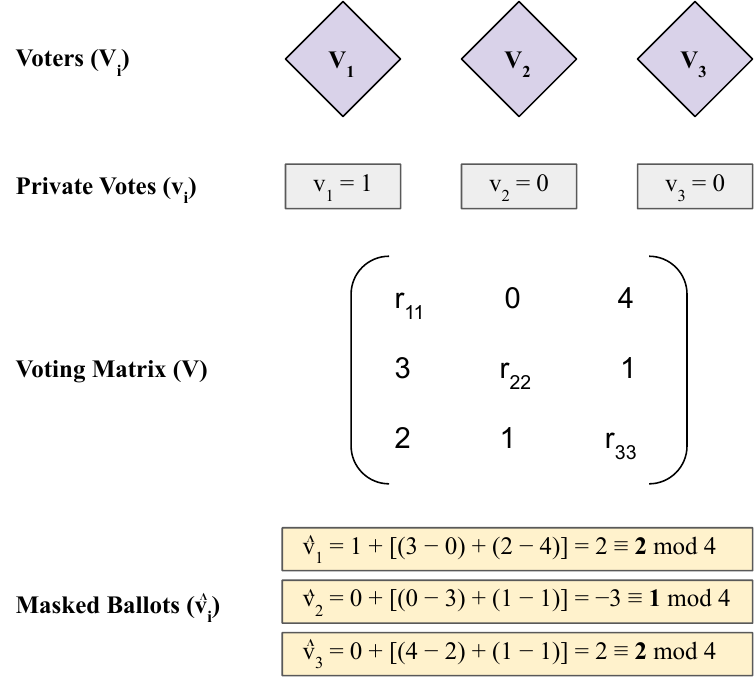}
\caption{An example of preparing a vote matrix $V$ and masked ballots $\hat{v}_i$ for $N=3$ voters.}
\label{fig:vote-mat-prep}
\end{figure}

  \begin{algorithm}[t]
  
    \DontPrintSemicolon
    \nonl \textbf{Commitment requires} $n$ $mod$ $4 = 0$, $m > 0$\;

    \For{Voter $V_{i} \in \mathcal{V}$}{
    {Every miner M$_j$ generates $m + 1$ balanced-uniform sequences (BUS) of $n$ qubits.}\;
    {\textbf{Communicate:} All BUS$_{M_{i}} \Rightarrow V_{i}$}\;
    {\textbf{Measurement:} V$_{i}$ randomly measures $m$ BUS from M$_{j}$}\;
    \If{Measurement \textbf{is not} balanced \textbf{and} uniform}{M$_{j}$ is cheating $\Rightarrow$ \textbf{Abort}}
    \textbf{Commitment:} V$_{i}$ operates on QS, i.e., the unrevealed-unmeasured BUS sequence: \quad \quad \quad \quad
    \textit{Commit} $00/01/10/11 \Rightarrow$ \textit{Apply} $R_X(0 / \frac{\pi}{2} / \pi / \frac{3\pi}{2})$\;
    {\textbf{$\forall$} commitments $V_{i} \Rightarrow M_{j}$: CS$_{V_{i}}^{M_{j}} \gets$ $n/2$ random bits}.\;
    {\textbf{Swapping:} QS$_{V_{i}}^{M_{j}}[2k-1]$ $\Leftrightarrow$ QS$_{V_{i}}^{M_{j}}[2k]$} $\forall$ CS$_{V_{i}}^{M_{j}}[k] \Rightarrow 1$\;
    {\textbf{Measurement:} Every miner M$_{j}$ measures $QS_{V_{i}}^{M_{j}}$ in $\{\ket{0}, \ket{1}\}$ basis, or in the $\{\ket{i}, \ket{-i}\}$ basis.}
    }
    \nonl\;
    \nonl\textbf{De-commitment requires} CS, QS\;
    
    \For{Voter $V_{i} \in \mathcal{V}$}{
        \textbf{Revelation}: \textbf{$\forall$} commitments $V_{i} \Rightarrow M_{j}$: Reveal $CS_{V_{i}}^{M_{j}}$
    }
    \For{Miner $M_{i} \in \mathcal{M}$}{
        {\textbf{Unswapping:} Perform the SWAPS from $CS_{V_{j}}^{M_{i}}$}\;
        {\textbf{Opening:} Miner M$_{i}$ begins restoring the $b_0b_1$ bits committed by Voter V$_{j}$}\;
        \textbf{Restore}{}\;
        {$b_0b_1 \Rightarrow 00$: $R_X(0)$(CBM state) = Original state}\;
        {$b_0b_1 \Rightarrow 01$: $R_X(\pi)$(CBM state) = Original state}\;
        {$b_0b_1 \Rightarrow 10$: $R_X(\pi/2)$(IBM state) = Original state }\;
        {$b_0b_1 \Rightarrow 11$: $R_X(3\pi/2)$(IBM state) = Original state }
    }
    \For{Voter $V_{i} \in \mathcal{V}$}{
        \textbf{Consensus}: Miners ($\mathcal{M}$) reach a consensus on the bit committed by V$_{i}$
    }
    \caption{Performing commitment and de-commitment of the masked ballot requiring two bits (k = 1), where (I)CBM state refers to the state yielded from (in)correct basis measurement.} \label{algo:algo-ballot-csqbc} 
   
    \end{algorithm}

\subsubsection{Ballot Preparation}
The first step for ballot commitment involves generating a voter matrix and distributing it via a quantum secure communication channel (QSCC) on a quantum blockchain. To build the integer voting matrix $\mathbf{V}$, each of the $N$ voters, V$_i$, would generate $N-1$ non-negative integers for $i^{th}$ row and the diagonal elements are then populated randomly with positive integers, such that the sum of each row is divisible by $N+1$.

Each voter V$_{i}$ uses a QSCC to share the element $V_{ij}$ of its row with the voter V$_{j}$. After that, each voter V$_{i}$ calculates their masked ballot $\hat{v}_i$ by computing the sum of all elements $V_{ji}$ received by them and adding to it a binary variable $v_{i}$ describing their vote. Finally, each V$_{i}$ commits their masked ballots to all the other miners M$_{j}$ in the blockchain via a commitment protocol discussed in the subsequent section, who will decide the result of voting by calculating the sum of the masked ballots. An example of the process among three voters is summarized in Fig. \ref{fig:vote-mat-prep}.


\subsubsection{Ballot Commitment}
The masked ballot is to be committed to every miner in the blockchain. This is achieved via cheat-sensitive quantum bit commitment (CSQBC) protocol, where every voter V$_{i}$ commits their masked ballot bitstring $\hat{v}_{i}$ to every miner M$_{j}$ by initiating the following two phases sequentially \cite{Sun2020}: 


\paragraph{Preparation phase}
 
In the preparation phase, every miner M$_j$ prepares $m + k$ balanced-uniform sequences (BUS) of $n$ qubits, where $n$ $mod$ $4 = 0$, $k = \lceil\log_2{(N)}\rceil$ and $m \geq k$. Each BUS is made of an equal number of qubits in the states $\ket{0}$, $\ket{1}$, $\ket{+i}$ and $\ket{-i}$ and is shuffled randomly before being sent to voter V$_i$. Each voter will choose $m$ sequences at random for the miner M$_j$ to reveal them and store the left-out $k$ sequences as $\text{QS}$. Then, voter V$_i$ measures these $m$ sequences of qubits in the relevant basis to check whether miner M$_j$ has prepared the qubits in the required form - qubits in states $\ket{0}$ / $\ket{1}$ and $\ket{+i}$ / $\ket{-i}$, in the $\hat{Z}$ and $\hat{Y}$ Pauli bases, respectively. Once the voter V$_i$ confirms miner M$_j$ has prepared all the communicated sequences uniformly, they proceed with the \textit{commitment phase}. Otherwise, they conclude that M$_j$ is cheating and abort the transaction.

\paragraph{Commitment phase}

In this phase, every voter V$_i$ commits their masked ballot $\hat{v}_i$ as \textit{two} bits ($b_0b_1$) at a time by applying quantum operators to each of the $k$ sequences in QS. More concretely, voter V$_i$ can commit $2k$ bits, where for each $b^k_0b^k_1 = 00/01/10/11$, they apply $R_X(\pi(b_0 + b_1 / 2))$ to corresponding k$^{th}$ QS and generates a classical bit string CS of length $n/2$ for shuffling it. They then send the $k$ permuted QS back to the Miner M$_j$ who would measure, who measures each qubit of the sequences in either of the $\hat{Z}$ and $\hat{Y}$ Pauli bases and stores the result.

\subsubsection{Ballot Decommitment}
Ballot tallying in the quantum binary voting is performed via de-commitment through the following two phases and compiled in the Algo. \ref{algo:algo-ballot-csqbc} after the commitment:

\paragraph{Opening phase}

Initially, every voter V$_{i}$ reveals their permutation sequences CS to every miner M$_{j}$ present in the blockchain to whom he had committed previously. Using each of the CS, miner M$_j$ decodes the original position of each qubit in the corresponding QS to gather swapping information, which is possible only because it is miner M$_j$ who had initially prepared it. This allows M$_j$ to determine whether each term in that QS was measured in the correct basis or not, and recover the bits committed by voter V$_i$ based on the state (I)CBM, i.e., the post-measurement state of all the qubits measured on a (in)correct basis, respectively. If applying $R_X(\pi(b_0 + b_1 / 2))$ to CBM (IBM) allows M$_j$ to obtain the original state, then the V$_{i}$ had committed either $00$ or $01$ ($10$ or $11$). This process is repeated until all $2k$ bits are recovered by each miner who will then attempt to reach a consensus on their values with all the other miners on the network.


\paragraph{Consensus phase}

All the miners $\in {M}$ run the quantum Byzantine agreement protocol to reach a consensus on the values of masked ballot $[\hat{v}_{1} \ldots \hat{v}_{2k}]$. To achieve the agreement, three miners are selected at a time $M_i$, $M_j$, and $M_k$ and share amongst them $T$ copies of a special three-qutrit state known as the Aharonov state $\ket{{A}} =  \frac{1}{\sqrt{6}}(\ket{0, 1, 2} + \ket{1, 2, 0} + \ket{2, 0, 1} - \ket{0, 2, 1} - \ket{1, 0, 2} -\ket{2, 1, 0})$, that can be mapped to a corresponding six-qubit state $\ket{{A}} = \frac{1}{\sqrt{6}} \big( \ket{00}_{M_1}(\ket{01}_{M_2}\ket{10}_{M_3} - \ket{10}_{M_2}\ket{01}_{M_3})$ $ +  \ket{01}_{M_1}(\ket{10}_{M_2}\ket{00}_{M_3} - \ket{00}_{M_2}\ket{10}_{M_3}) + \ket{10}_{M_1}(\ket{00}_{M_2}\ket{01}_{M_3} - \ket{01}_{M_2}\ket{00}_{M_3}) \big)$.


Furthermore, to reach a consensus on the value of the masked ballot, the protocol is run sequentially for each bit in masked ballot $\hat{v}_{i}$. In every iteration, a miner M$_{i}$ is elected as a leader at random, who sends the bit $x$ in their possession to the other two miners $\{M_j, M_k\}$ classically and measures their pair of shared qubits in the computation basis for every $\ket{A}$ state until their measurements equal $x$. They send the index $t$ of that state to both M$_j$ and M$_k$, who store the collective information as \{$x_j$, $I_j$\} and \{$x_k$, $I_k$\}, respectively. Then, both M$_{j}$ and $M_{k}$ check the consistency of their results by first cross-checking them with the copy of the masked ballot bitstring $\hat{v}_{i}$ they possess and then by measuring their pairs of qubits in the computation basis. They set their consistency flag if they have consistent data, i.e., all the indices in $I_{j(k)}\neq x_{j(k)}$, or else they will put it down. Finally, they communicate their consistency flags with each other and reach a consensus if their flags are in agreement. However, if their flags disagree, they would know that M$_i$ was complicit and hence update their value of the bit $x$ from the other miner's value. 


\begin{figure*}[!ht]
\centering
\begin{subfigure}{0.6\linewidth}
\includegraphics[width=\linewidth]{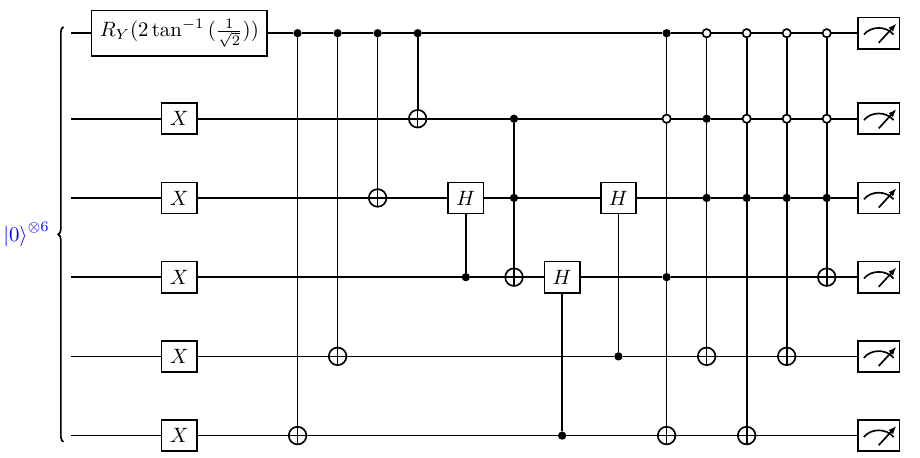} 
\caption{}
\end{subfigure}\hfill
\begin{subfigure}{0.4\linewidth}
\includegraphics[width=\linewidth]{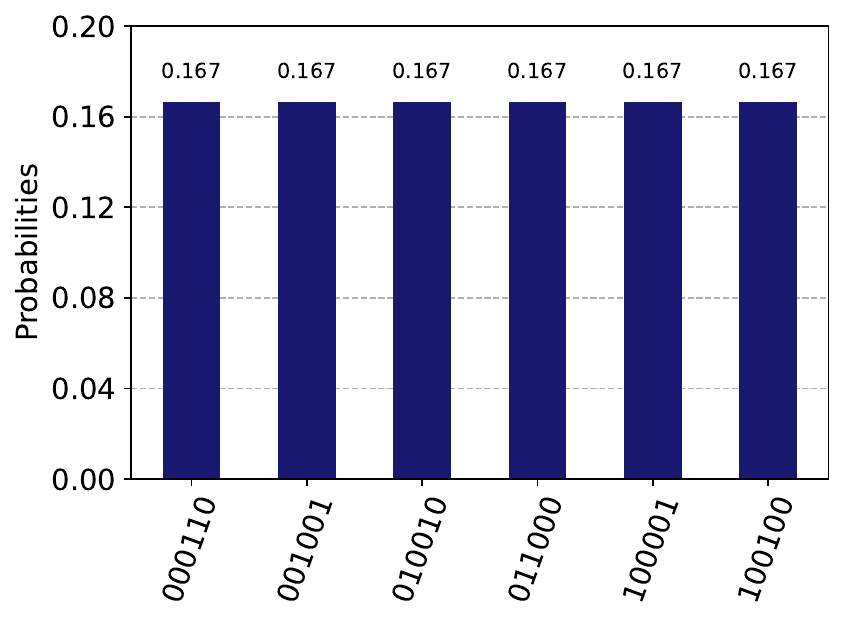} 
\caption{}
\end{subfigure}\hfill
\caption{(a) State preparation circuit for the Aharonov state $\ket{A}$. When decomposed into the basis gates $[U, \text{CNOT}]$ supported by IBM quantum hardware, the depth of the circuit would be $146$, and it would use $97$ single-qubit $U$ gates and $83$ two-qubit $\text{CNOT}$ gates. (b) The probability distribution of the state prepared by the circuit on a noiseless device.}
\end{figure*}

\begin{algorithm}[!t]

\DontPrintSemicolon
\nonl\textbf{Agreement requires} $N > 0$, $\mathcal{M} \neq \{\phi\}$, $\hat{v}_{i}$\;
\For{t $\gets$ 1, T}{
    {Prepare $\ket{\mathcal{A}}^{t}$}\;
    {\textbf{Communicate:}  Distribute $\ket{\mathcal{A}}^{t}_{M_{j}}$ to each $M_{j} \in \mathcal{M}$}
}

\For{bit $\in \hat{v}_{i}$}{
    {Select the leader M$_{j}$}\;
    {$x$ $\gets$ M$_{j}$'s \textbf{bit}}\;
    {\textbf{Communicate:} $\{$M$_{k},\ $M$_{l}\}$  $\gets$ $x$}\;
    {\textbf{Measurement:} Outcomes $\gets$ All $\ket{\mathcal{A}}^{t}_{M_{j}}$ are
    measured by M$_{j}$}\;
    \For{\text{Outcomes} $O_{i}$}{
        \If{$O_{t} ==  x$}{
            \If{M$_{j}$ is \textbf{Honest}}{
                {\textbf{Communicate:} M$_{k}: \{x_{k}, I_{k}\}$  $\gets$ $x$, $t$}\;
                {\textbf{Communicate:} M$_{l}: \{x_{l}, I_{l}\}$  $\gets$ $x$, $t$}\;
            \Else{
                {\textbf{Communicate:} M$_{k}: \{x_{k}, I_{k}\}$  $\gets$ $x$, $t$}\;
                {\textbf{Communicate:} M$_{l}: \{x_{l}, I_{l}\}$  $\gets$ $1-x$, $t$}}
            }
        }
    }

\For{$\{x_{k}, I_{k}\}$ and $\{x_{l}, I_{l}\}$}{    
    \If{$x_{p} == M_{p}$'s bit}{
        {\textbf{Measurement:} y $\gets \ket{\mathcal{A}}^{t}_{M_{p}}$ measured by $M_{p}$}\;
        \If{all $y\neq x_{p}$}{
            {$Flag_{M_{p}}$ $\gets$ \textbf{Consistent}}\;
        \Else
            {$Flag_{M_{p}}$ $\gets$ \textbf{Inconsistent}}
        }
    }
}

{\textbf{Communicate:} $Flag_{M_{k}} \Leftrightarrow Flag_{M_{l}}$}\;

\If{$Flag_{M_{k}} == Flag_{M_{l}}$ \textbf{and} $x_{k} == x_{l}$}{
    {\textbf{Agreement:} \textit{Successful Broadcast}}\;
\ElseIf{$Flag_{M_{k}} != Flag_{M_{l}}$}{
    {\textbf{Communicate:} Update to \textbf{consistent} bit value}\;
    {\textbf{Agreement:} \textit{Successful Broadcast}}}
\Else{
    {\textbf{Convince}: M$_{k} \Rightarrow$ M$_{l}$}}
    \If{Convinced}{
        {\textbf{Agreement:} \textit{Successful Broadcast}}\;
    \Else{
        {\textbf{Disagreement:} \textit{Detectable Broadcast}}}
    }
}

}
\caption{Performs Byzantine agreement among miners ($\mathcal{M}$) for a masked ballot $v_{i}$ using $N$ Aharonov states $\ket{\mathcal{A}}$.} \label{algo:algo-bap}

\end{algorithm}

Additionally, there could be a case where both miners M$_j$ and M$_k$ have consistent data but different values of the bit. In this scenario, miner M$_j$ tries to convince M$_k$ that they are not complicit, i.e., their data is trustworthy, by sending all indices $t \in I_{j}$ with the opposite answer. Miner M$_k$ then checks whether he got \textit{enough} indices $t$, such that all $t \notin I_k$ and post-measurement state of their qubits of $\ket{A}$ are predominantly $\ket{10}$. If they succeed, the data and the ballot bit are updated for the other miner. It is important to note here the term \textit{enough} can be represented by an agreement probability $\lambda$, i.e., the mean fraction of convincing indices should be $\geq\lambda$. This leads to two variations of successful terminations of the agreement protocol. The first kind, where there is a disagreement at the end of the convincing step is referred to as \textit{detectable} broadcasts and the second kind, where an agreement is reached is called \textit{successful} broadcast. We present the whole procedure in the Algo. \ref{algo:algo-bap} and the success probability $P(S)$ of both variations with the number of $\ket{{A}}$ state miners $\in {M}$ will possess is shown in Fig. \ref{fig:fig6}.

\section{Security Analysis}\label{QVP:Sec5}


In this section, we present an analysis of the security of the proposed quantum binary voting protocol against internal (i.e., the complacency of voters and miners) and external attacks (i.e., by some unwarranted adversary, might affect/tamper with the final results). Furthermore, we estimate the required computational resources for these attacks and show how the CSQBC and quantum blockchain ensure that the protocol satisfies the blockchain requirements like binding, transparency, fairness, etc., as discussed in Section \ref{VRP}.

\subsection{Quantum Bit Commitment}\label{QVP:Sec5:QBC}

CSQBC protocol \cite{bib_Sun} showed that for Alice to commit every two bits to Bob, the amount of resource of bits and qubits needed is $(m  + 1)\times n$, where $m$ is the number of sequences and $n$ is the length of each sequence. For Alice to successfully cheat, she must change a committed bit before revealing it to Bob. The successful probability $p_a$ for this to happen is $(1/2)^{n}$, and there’s a probability of $1/2$ for her to get detected. In contrast, Bob cheats successfully if he gets some knowledge of the committed bit before Alice reveals it to him. The success probability $p_b$ for this to happen is $(1 - 1/4^n) / (m + 1) \approx 1/(m + 1)$ for large $n$, and there’s a possibility for more than $1/2$ for him to get detected. This enhanced security of the CSQBC makes the protocol binding and ensures fairness.

\subsection{Internal Attacks}
Quantum bit commitment ensures that neither voters can change their masked ballots nor miners can get non-trivial information regarding them before the tally phase. However, there still can be complacency in how voters and miners behave. The first way would be for a voter V$_{i}$ to send the wrong values of V$_{ij}$ to V$_{j}$. Due to access to the quantum secure communication channel, other voters will not know if they received the correct value and construct the voter matrix $\mathbf{V}$. However, nodes on the blockchain that are not participating in the election process can act as independent auditors. They can inquire about the values V$_{ij}$ (V$_{ji}$) from both V$_{i}$ and V$_{j}$, and catch any cheating amongst the voters. Another possibility for an internal attack would be for one of the miners to be complicit and disrupt the consensus phase. As discussed in Section \ref{QVP:Sec7}, given sufficient quantum resources while complicit miners are $\leq ||M||/3$, the consensus probability $p_c$ would remain almost equivalent to $1$. Moreover, since the quantum blockchain is transparent by design, it ensures users self-tally simply by calculating the sum of the masked ballots and verifying whether the correct masked ballot is successfully uploaded to the blockchain database.
  
\subsection{External Attacks}
In a voting protocol, an external adversary would prefer to manipulate or tally the ballots before the tallying phase. For an adversary to do so, it can perform a CNOT or Ping-Pong attack on the quantum blockchain \cite{Li2013}. The adversary would require the same number of qubits as present with the nodes acting as miners in the blockchain to perform such an attack which will be infeasible in large blockchain networks. Moreover, an adversary will not be able to manipulate the ballots before the opening phase due to the concealing and binding property of CSQBC. Even if it tries to manipulate them during the consensus phase, it becomes equivalent to an internal attack, for which the quantum Byzantine protocol is sufficiently resilient as discussed before. Another way a CNOT or Ping-Pong attack could be done is by getting hold of the key used for authentication on the quantum blockchain using the QKD protocol \cite{bib_qkd}. However, even if the adversary gets the key, it cannot append another ballot using voter’s authentication as quantum blockchains don't allow double-spending attacks. Accordingly, by its design, the quantum blockchain maintains eligibility, i.e., only authenticated voters can successfully communicate with the miners.

\begin{figure*}[!ht]
\centering
\begin{subfigure}{0.48\linewidth}
\includegraphics[width=\linewidth]{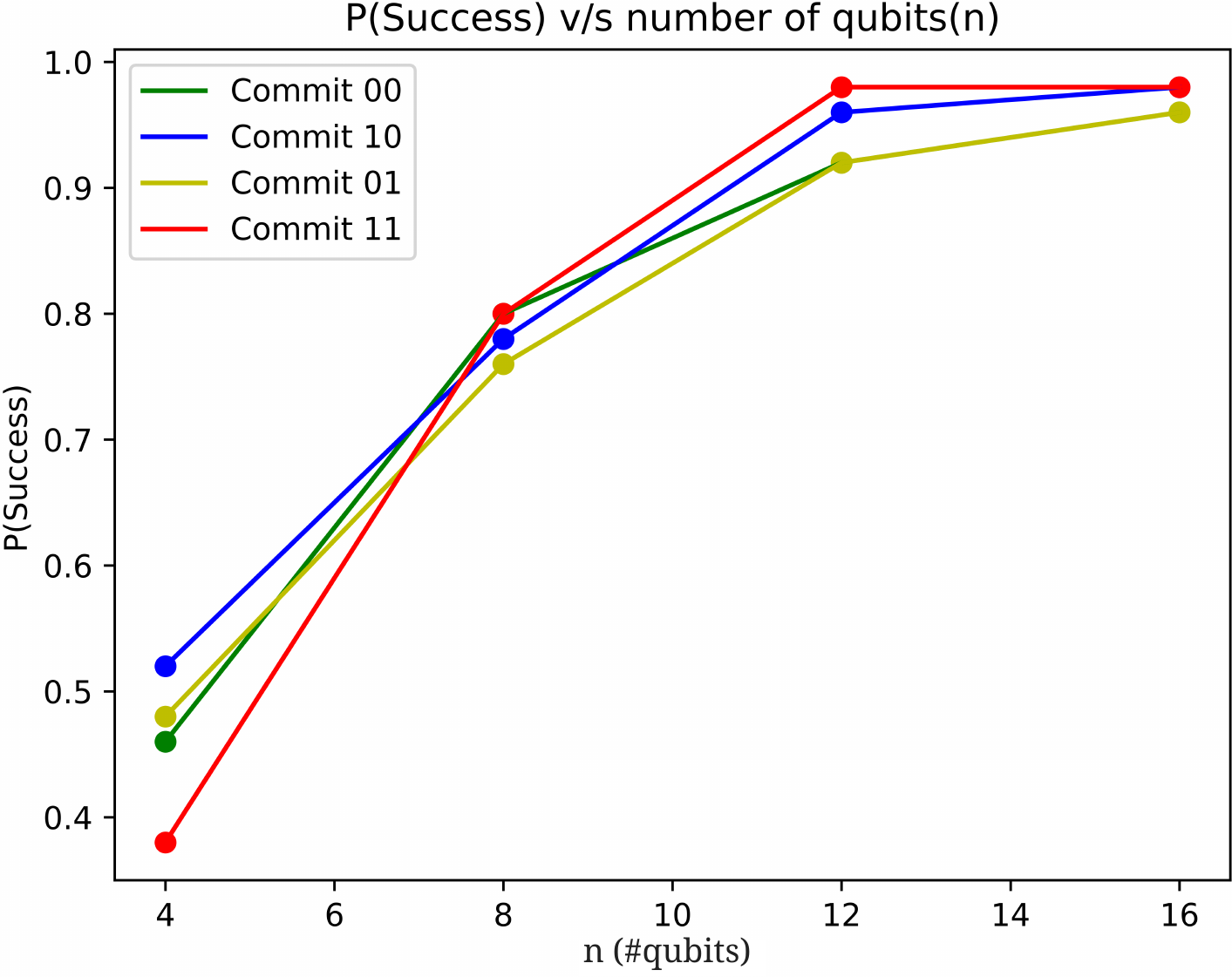} 
\caption{}
\label{fig:csqbc_fig4a}
\end{subfigure}\hfill
\begin{subfigure}{0.48\linewidth}
\includegraphics[width=\linewidth]{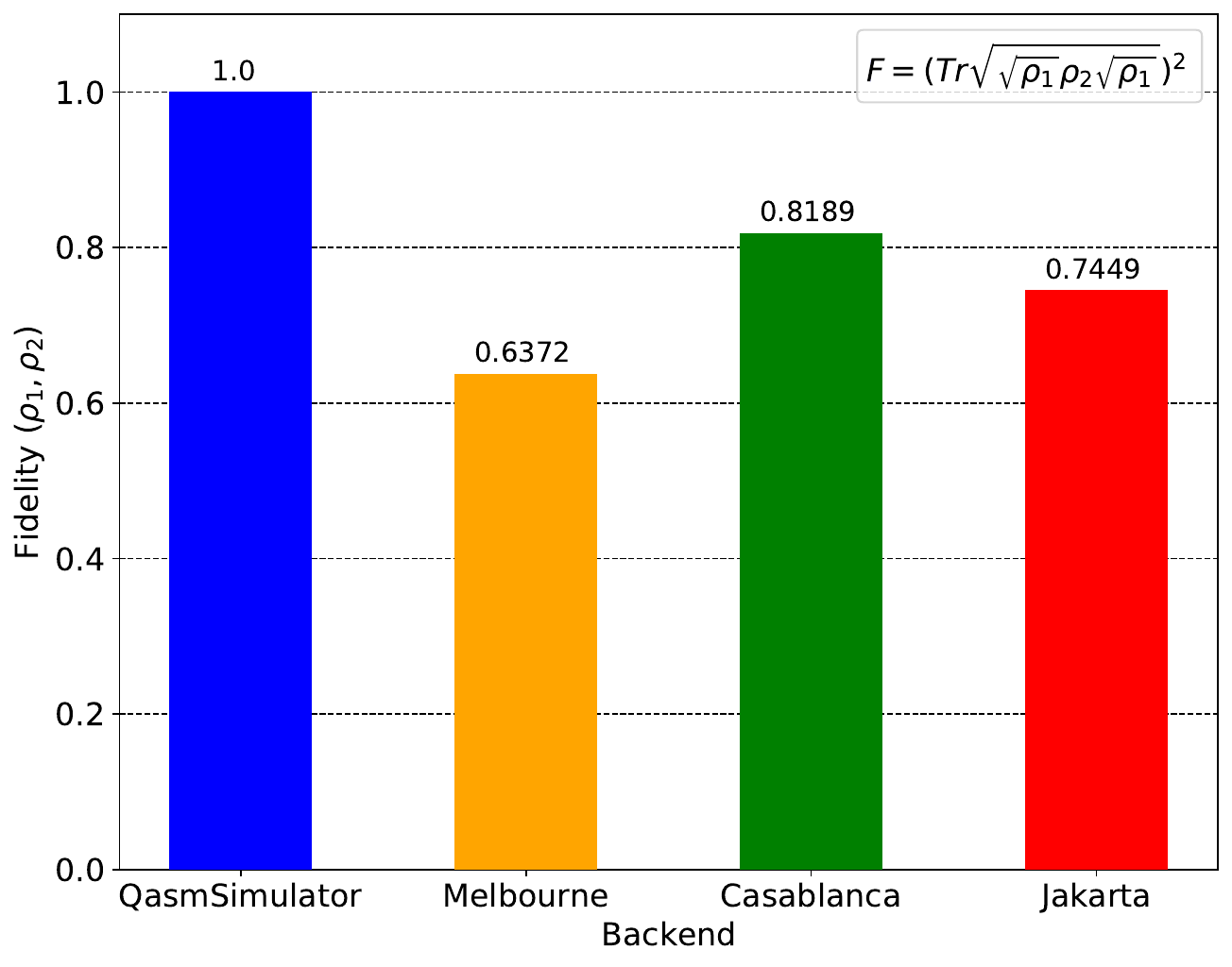}
\caption{}
\label{fig:qba_fig4b}
\end{subfigure}\hfill
\caption{(a) Probability of success for committing the right bit via CSQBC protocol is plotted against $n$, i.e., the number of qubits required to generate the sequence QS, for different bitstrings. (b) Fidelity of $\ket{{A}}$ states prepared in different IBMQ backends with respect to $\ket{{A}}$ state prepared via an ideal noise-less simulation. Here, we have used noise models from different hardware backends for our simulations.}
\end{figure*}

\section{Simulations and Results} \label{QVP:Sec6}

In the following subsections, we show the success probability of commitment of two-bit masked ballots and achieving consensus over their values with three miners using the cheat cheat-sensitive quantum bit commitment (CSQBC) and quantum Byzantine agreement (QBA) protocols, respectively. Both protocols have been implemented via IBM Qiskit \cite{bib_IBM} and the working of the former has been additionally verified with a subsequent implementation using SimulaQron \cite{bib_simqron}.

\subsection{Commitment protocol}

The success probability for the CSQBC scheme is defined as the number of times bits committed by the voters V$_i$ in the commitment phase are the same as the miners M$_j$ after the opening phase when the simulations are performed multiple times. We use this as a metric and repeat the scheme $1000$ times for $n \in [4,8,12,16]$ using the SimulaQron library to perform two bits. The results are shown in Fig. \ref{fig:csqbc_fig4a}, where the failures in committing the bit occurred due to the intrinsic probabilistic nature of the protocol and not due to noise as the simulations are performed in an ideal environment. Moreover, the success rate does not depend on the value of $m$, as voter V$_i$ and miner M$_j$ work primarily only with one qubit sequence, $QS$. We also performed a similar simulation on the IBM Quantum platform with $n = 4, m = 1, k = 1$, where we achieved similar results. In this case, as the entire commitment from a voter V$_{i}$ to a miner M$_{j}$ will take $32$ qubits, the simulations were done in a partwise manner. The swap operators were used for transferring qubits (encoding bit data) between miners and voters, and rotation gates were utilized in creating the required basis for measurements.

\subsection{Consensus Protocol}

\begin{figure*}[!ht]
\centering
\begin{subfigure}{0.5\linewidth}
\includegraphics[width=\linewidth]{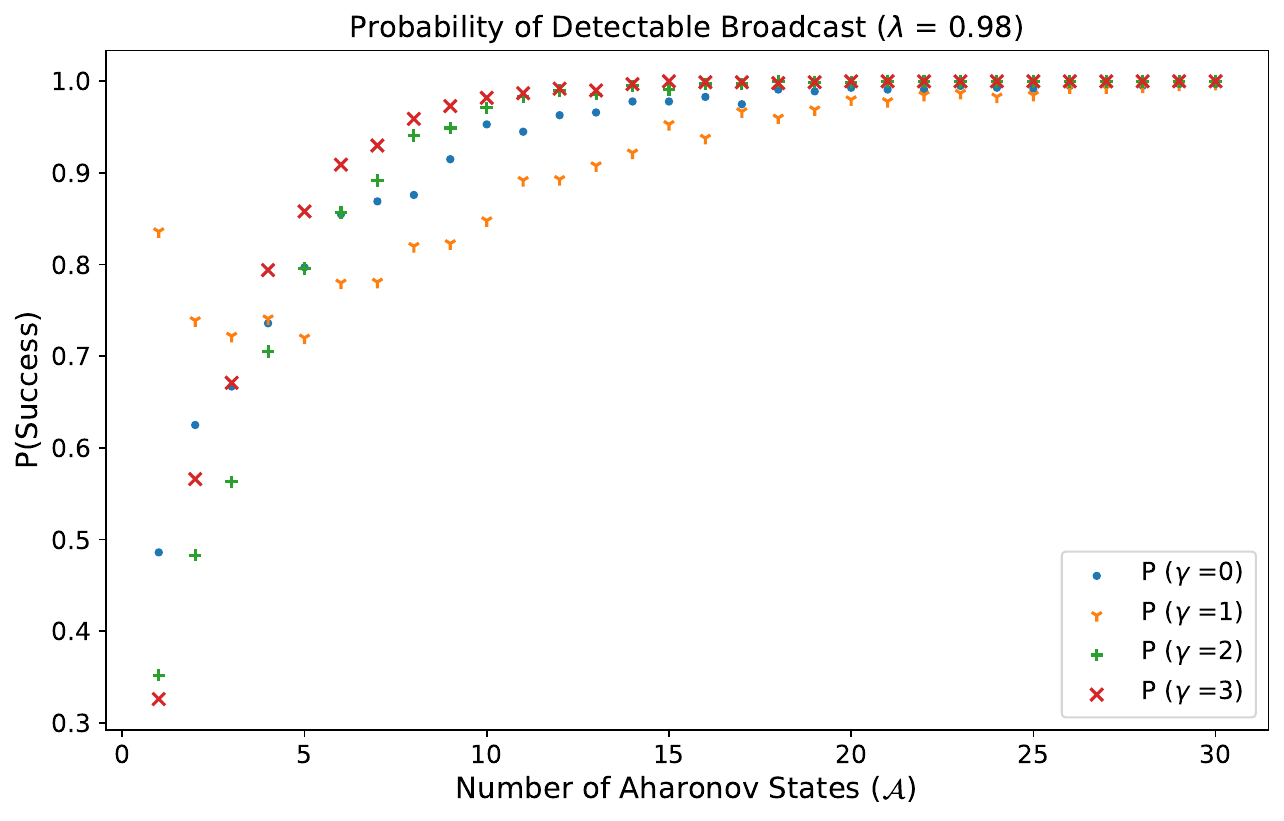} 
\caption{}
\end{subfigure}\hfill
\begin{subfigure}{0.5\linewidth}
\includegraphics[width=\linewidth]{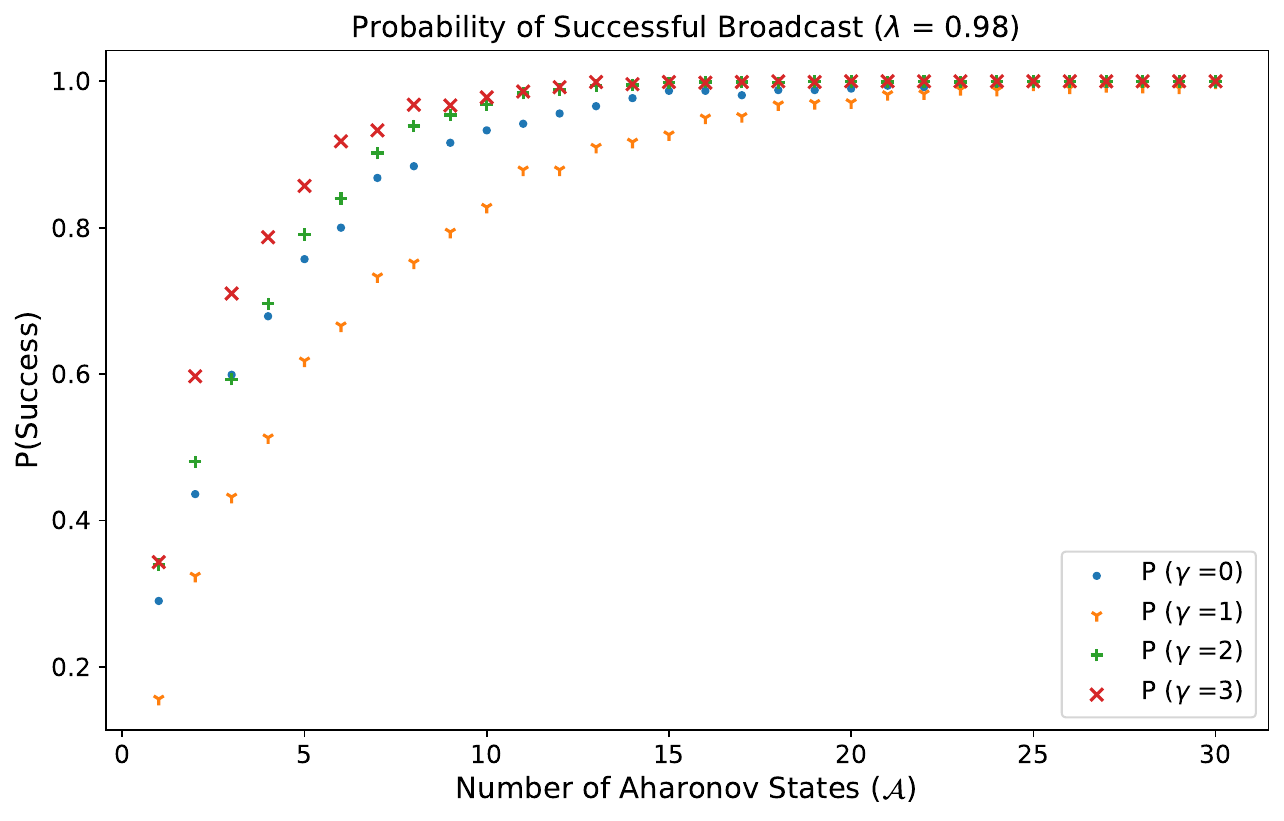} 
\caption{}
\end{subfigure}\hfill
\caption{The probability of (a) detectable and (b) successful broadcasts increases with an increment in the number of Aharonov states $\ket{{A}}$ that have been distributed within the miners ${M}$, even if there are complicit miner(s), for any sufficient agreement probability $\lambda\geq0.9$. Here, the parameter $\gamma$ represents that complicit behaviour exhibited in the miner within the blockchain system in any iteration, where: (i) $\gamma=0$, any of the three miners is complicit at random and (ii) $\gamma =i\neq0$, only miner M$_{i}$ is always complicit.}
\label{fig:fig6}
\end{figure*}

 The QBA protocol involves distributing $T$ copies of the Aharonov $\ket{{A}}$ amongst all miners. The preparation of one $\ket{{A}}$ on different IBM quantum devices (\ref{fig:qba_fig4b}) is performed and then their construction by these devices is assessed using the quantum state tomography, where given the measurement data $\{m_{i}\}$, the maximum likelihood state is found by minimizing the log function, ${L}_{\text{log}} \sum_{i} [m_{i}-\text{Tr}(\sigma_{i}\rho)]^{2}$ for $\rho \geq 0$ and $\text{Tr}(\rho) = 1$. In our experiments, we considered $\rho$ to be the experimental noisy density matrix of the maximum likelihood state. We used it to calculate the fidelity ${F}(\rho, \rho_0) = \left[\operatorname {Tr} {\sqrt {{\sqrt {\rho_{\text{noisy}} }}\rho_{\text{ideal}}{\sqrt {\rho_{\text{noisy}} }}}}\right]^{2}$ with respect to the $\rho_{0}$, the density matrix for the ideal $\ket{{A}}$. From the results, it is obvious that out of all three hardware backends that were available publically at the time of running these simulations, only Casablanca could do a decent state preparation that can be improved further using error-mitigation routines. Furthermore, the $P (\text{success})$, i.e., the success probability with different numbers of $\ket{{A}}$ states are obtained in the presence of at most one complicit adversary present amongst the miners ${M}$ and is shown in Fig. \ref{fig:fig6}. It is noted that for $n>25$ $\ket{{A}}$ states, the success probability for both detectable and successful broadcasts grew to $\approx100$ percent. Since each $\ket{{A}}$ requires $6$ qubits, this means the qubit requirement would be $\approx 200$ qubits for each consensus iteration, unless qubits could be reused and reshared among the miners.

\section{Discussion and Conclusion}\label{QVP:Sec7}

This paper demonstrates the implementation of a novel voting protocol based on quantum blockchain that satisfies requirements regarding anonymity, receipt-freeness, fairness, eligibility, non-reusability, binding, self-tallying, and verifiability. Our proposed protocol improves upon the one given by \cite{bib_Sun} in the ballot preparation and commitment step. We use $n \times (n-1)$ non-negative integers for building the voting matrix, skipping the diagonal element, making it possible to verify an internal attack in the quantum blockchain by allowing the non-participating nodes to act as an independent auditor. Additionally, we perform a bit commitment protocol that is implementable by current hardware as it doesn't rely on resource-intensive quantum memory and multipartite entanglement. We also show a consensus mechanism based on Byzantine agreement protocol that could be implemented efficiently via qubit reusing strategies. 

By implementing these protocols for the simulations, we could compare the IBM Quantum (IBMQ) and SimulaQron platforms for performing quantum simulations. IBMQ provides access to hardware for executing low-depth quantum circuits and the Qiskit software stack for building arbitrary quantum circuits. By contrast, SimulaQron is more suitable for classically simulating quantum networking protocols and utilizes the concept of virtual qubits to execute decentralized algorithms with reduced overheads. However, due to the limitations of both these platforms regarding functional quantum secure communication channels, security and implementation burdens still need to be overcome to achieve a fully implementable quantum blockchain and demonstrate an IoT network integrated with quantum bit commitment and Byzantine agreement protocols. We leave this as an extension of this work for the future.

Based on our work and state of the current literature \cite{Liu2024, Wazid2024, Jain2023}, we conclude that the IoT framework can benefit from adopting the quantum blockchain technology. In this spirit, we provide an effective and efficient voting protocol that would enable tasks requiring consensus, pushing forward the feasibility of the quantum blockchain for this purpose on the current hardware and steering the effort to come up with simpler solutions to the individual components that were previously thought to be cumbersome and complicated.

\section*{Acknowledgments}
\label{qlock_acknowledgments}
U.~A. acknowledges the assistance received from S.I~Aadharsh~Raj for implementing the cheat-sensitive quantum bit commitment scheme and Zoltán Guba for implementing the Byzantine agreement protocol.




\bibliographystyle{IEEEtran}
%

\bibliography{qvote}

\begin{thebibliography}{10}
\providecommand{\url}[1]{#1}
\csname url@samestyle\endcsname
\providecommand{\newblock}{\relax}
\providecommand{\bibinfo}[2]{#2}
\providecommand{\BIBentrySTDinterwordspacing}{\spaceskip=0pt\relax}
\providecommand{\BIBentryALTinterwordstretchfactor}{4}
\providecommand{\BIBentryALTinterwordspacing}{\spaceskip=\fontdimen2\font plus
\BIBentryALTinterwordstretchfactor\fontdimen3\font minus \fontdimen4\font\relax}
\providecommand{\BIBforeignlanguage}[2]{{%
\expandafter\ifx\csname l@#1\endcsname\relax
\typeout{** WARNING: IEEEtran.bst: No hyphenation pattern has been}%
\typeout{** loaded for the language `#1'. Using the pattern for}%
\typeout{** the default language instead.}%
\else
\language=\csname l@#1\endcsname
\fi
#2}}
\providecommand{\BIBdecl}{\relax}
\BIBdecl

\bibitem{5gBC}
\BIBentryALTinterwordspacing
A.~Sharma, M.~Singh, M.~Gupta, N.~Sukhija, and P.~K. Aggarwal, ``{IoT and blockchain technology in 5G smart healthcare},'' \emph{Blockchain Applications for Healthcare Informatics}, pp. 137--161, 2022. [Online]. Available: \url{https://doi.org/10.1016/B978-0-323-90615-9.00004-9}
\BIBentrySTDinterwordspacing

\bibitem{bib_Hardwick}
\BIBentryALTinterwordspacing
F.~{Sheer Hardwick}, A.~{Gioulis}, R.~{Naeem Akram}, and K.~{Markantonakis}, ``{E-Voting with Blockchain: An E-Voting Protocol with Decentralisation and Voter Privacy},'' \emph{arXiv e-prints}, may 2018. [Online]. Available: \url{https://doi.org/10.48550/arXiv.1805.10258}
\BIBentrySTDinterwordspacing

\bibitem{Sriman2020}
\BIBentryALTinterwordspacing
B.~Sriman, S.~Ganesh~Kumar, and P.~Shamili, \emph{{Blockchain Technology: Consensus Protocol Proof of Work and Proof of Stake}}.\hskip 1em plus 0.5em minus 0.4em\relax Springer Singapore, sep 2020, p. 395–406. [Online]. Available: \url{http://dx.doi.org/10.1007/978-981-15-5566-4_34}
\BIBentrySTDinterwordspacing

\bibitem{Gidney2021}
\BIBentryALTinterwordspacing
C.~Gidney and M.~Ekerå, ``{How to factor 2048 bit RSA integers in 8 hours using 20 million noisy qubits},'' \emph{Quantum}, vol.~5, p. 433, apr 2021. [Online]. Available: \url{http://dx.doi.org/10.22331/q-2021-04-15-433}
\BIBentrySTDinterwordspacing

\bibitem{Bova2021}
\BIBentryALTinterwordspacing
F.~Bova, A.~Goldfarb, and R.~G. Melko, ``{Commercial applications of quantum computing},'' \emph{EPJ Quantum Technology}, vol.~8, no.~1, jan 2021. [Online]. Available: \url{http://dx.doi.org/10.1140/epjqt/s40507-021-00091-1}
\BIBentrySTDinterwordspacing

\bibitem{Wang2022}
\BIBentryALTinterwordspacing
W.~Wang, Y.~Yu, and L.~Du, ``{Quantum blockchain based on asymmetric quantum encryption and a stake vote consensus algorithm},'' \emph{Scientific Reports}, vol.~12, no.~1, may 2022. [Online]. Available: \url{http://dx.doi.org/10.1038/s41598-022-12412-0}
\BIBentrySTDinterwordspacing

\bibitem{S2024}
\BIBentryALTinterwordspacing
M.~G. S, C.~Mulay, K.~Durai, G.~Murali, J.~A. Ibrahim Syed~Masood, V.~Vijayarajan, K.~Gautam, N.~S. Kalyan~Chakravarthy, S.~Suresh~Kumar, S.~Agarwal, M.~S, V.~V, D.~Asirvatham, S.~Brohi, C.~V. C, and A.~S, ``{Quantum blockchain: Trends, technologies, and future directions},'' \emph{IET Quantum Communication}, vol.~5, no.~4, p. 516–542, dec 2024. [Online]. Available: \url{http://dx.doi.org/10.1049/qtc2.12119}
\BIBentrySTDinterwordspacing

\bibitem{bib_Sun}
\BIBentryALTinterwordspacing
X.~Sun, Q.~Wang, P.~Kulicki, and M.~Sopek, ``{A Simple Voting Protocol on Quantum Blockchain},'' \emph{International Journal of Theoretical Physics}, vol.~58, no.~1, pp. 275--281, oct 2018. [Online]. Available: \url{https://doi.org/10.1007/s10773-018-3929-6}
\BIBentrySTDinterwordspacing

\bibitem{bib_IBM}
\BIBentryALTinterwordspacing
G.~Aleksandrowicz, T.~Alexander, P.~Barkoutsos, L.~Bello, Y.~Ben-Haim, D.~Bucher, F.~J. Cabrera-Hernández, and e.~a. Carballo-Franquis, ``\BIBforeignlanguage{en}{{Qiskit: An Open-source Framework for Quantum Computing}},'' 2019. [Online]. Available: \url{https://zenodo.org/record/2562110}
\BIBentrySTDinterwordspacing

\bibitem{bib_simqron}
\BIBentryALTinterwordspacing
A.~Dahlberg and S.~Wehner, ``{{SimulaQron}{\textemdash}a simulator for developing quantum internet software},'' \emph{Quantum Science and Technology}, vol.~4, no.~1, p. 015001, sep 2018. [Online]. Available: \url{https://doi.org/10.1088/2058-9565/aad56e}
\BIBentrySTDinterwordspacing

\bibitem{Mocnej2018}
\BIBentryALTinterwordspacing
J.~Mocnej, W.~K. Seah, A.~Pekar, and I.~Zolotova, ``{Decentralised IoT Architecture for Efficient Resources Utilisation},'' \emph{IFAC-PapersOnLine}, vol.~51, no.~6, p. 168–173, 2018. [Online]. Available: \url{http://dx.doi.org/10.1016/j.ifacol.2018.07.148}
\BIBentrySTDinterwordspacing

\bibitem{Tripathi2023}
\BIBentryALTinterwordspacing
G.~Tripathi, M.~A. Ahad, and G.~Casalino, ``{A comprehensive review of blockchain technology: Underlying principles and historical background with future challenges},'' \emph{Decision Analytics Journal}, vol.~9, p. 100344, dec 2023. [Online]. Available: \url{http://dx.doi.org/10.1016/j.dajour.2023.100344}
\BIBentrySTDinterwordspacing

\bibitem{9781107002173}
\BIBentryALTinterwordspacing
M.~A. Nielsen and I.~L. Chuang, \emph{{Quantum Computation and Quantum Information: 10th Anniversary Edition}}.\hskip 1em plus 0.5em minus 0.4em\relax Cambridge University Press, 2011. [Online]. Available: \url{https://www.amazon.com/Quantum-Computation-Information-10th-Anniversary/dp/1107002176?}
\BIBentrySTDinterwordspacing

\bibitem{Kho2022}
\BIBentryALTinterwordspacing
Y.-X. Kho, S.-H. Heng, and J.-J. Chin, ``{A Review of Cryptographic Electronic Voting},'' \emph{Symmetry}, vol.~14, no.~5, p. 858, apr 2022. [Online]. Available: \url{http://dx.doi.org/10.3390/sym14050858}
\BIBentrySTDinterwordspacing

\bibitem{bib_Cono}
\BIBentryALTinterwordspacing
M.~Conoscenti, A.~Vetrò, and J.~C. De~Martin, ``{Blockchain for the Internet of Things: A systematic literature review},'' in \emph{2016 IEEE/ACS 13th International Conference of Computer Systems and Applications (AICCSA)}, 2016, pp. 1--6. [Online]. Available: \url{https://doi.org/10.1109/AICCSA.2016.7945805}
\BIBentrySTDinterwordspacing

\bibitem{bib_Novo}
\BIBentryALTinterwordspacing
O.~Novo, ``{Blockchain Meets {IoT}: An Architecture for Scalable Access Management in {IoT}},'' \emph{{IEEE} Internet of Things Journal}, vol.~5, no.~2, pp. 1184--1195, apr 2018. [Online]. Available: \url{https://doi.org/10.1109/jiot.2018.2812239}
\BIBentrySTDinterwordspacing

\bibitem{bib_Reyna}
\BIBentryALTinterwordspacing
A.~Reyna, C.~Mart{\'{\i}}n, J.~Chen, E.~Soler, and M.~D{\'{\i}}az, ``{On blockchain and its integration with {IoT}. Challenges and opportunities},'' \emph{Future Generation Computer Systems}, vol.~88, pp. 173--190, nov 2018. [Online]. Available: \url{https://doi.org/10.1016/j.future.2018.05.046}
\BIBentrySTDinterwordspacing

\bibitem{bib_Kouicem}
\BIBentryALTinterwordspacing
D.~E. Kouicem, A.~Bouabdallah, and H.~Lakhlef, ``{Internet of things security: A top-down survey},'' \emph{Computer Networks}, vol. 141, pp. 199--221, aug 2018. [Online]. Available: \url{https://doi.org/10.1016/j.comnet.2018.03.012}
\BIBentrySTDinterwordspacing

\bibitem{bib_Andrew}
\BIBentryALTinterwordspacing
A.~Lewis, \emph{{The basics of bitcoins and blockchains: an introduction to cryptocurrencies and the technology that powers them}}.\hskip 1em plus 0.5em minus 0.4em\relax Coral Gables, FL: Mango Publishing, 2021. [Online]. Available: \url{https://doi.org/10.1007/s11408-020-00374-0}
\BIBentrySTDinterwordspacing

\bibitem{bib_Liu}
\BIBentryALTinterwordspacing
Y.~Liu and Q.~Wang, ``{An E-voting Protocol Based on Blockchain},'' \emph{IACR Cryptol. ePrint Arch.}, vol. 2017, p. 1043, 2017. [Online]. Available: \url{https://eprint.iacr.org/2017/1043}
\BIBentrySTDinterwordspacing

\bibitem{bib_V1}
\BIBentryALTinterwordspacing
R.~Rojas, \emph{{Encyclopedia of computers and computer history}}.\hskip 1em plus 0.5em minus 0.4em\relax Chicago: Fitzroy Dearborn, 2001. [Online]. Available: \url{https://search.worldcat.org/en/title/606796819}
\BIBentrySTDinterwordspacing

\bibitem{bib_V2}
\BIBentryALTinterwordspacing
D.~A. Kumar and T.~U.~S. Begum, ``{Electronic voting machine -- A review},'' in \emph{International Conference on Pattern Recognition, Informatics and Medical Engineering (PRIME-2012)}, 2012, pp. 41--48. [Online]. Available: \url{https://doi.org/10.1109/ICPRIME.2012.6208285}
\BIBentrySTDinterwordspacing

\bibitem{bib_prev4}
\BIBentryALTinterwordspacing
W.~Yu-Wu, W.~Xiang-He, and Z.~Zhao-Hui, ``{Quantum voting protocols based on the non-symmetric quantum channel with controlled quantum operation teleportation},'' \emph{Acta Physica Sinica}, vol.~62, no.~16, p. 160302, 2013. [Online]. Available: \url{https://doi.org/10.7498/aps.62.160302}
\BIBentrySTDinterwordspacing

\bibitem{bib_commit}
\BIBentryALTinterwordspacing
X.~Sun, F.~He, and Q.~Wang, ``{Impossibility of Quantum Bit Commitment, a Categorical Perspective},'' \emph{Axioms}, vol.~9, no.~1, p.~28, mar 2020. [Online]. Available: \url{https://doi.org/10.3390/axioms9010028}
\BIBentrySTDinterwordspacing

\bibitem{Sun2020}
\BIBentryALTinterwordspacing
X.~Sun, Q.~Wang, and F.~He, ``{An Arbitrarily Concealing and Practically Binding Quantum Bit Commitment Protocol},'' \emph{International Journal of Theoretical Physics}, vol.~59, no.~11, p. 3464–3475, oct 2020. [Online]. Available: \url{http://dx.doi.org/10.1007/s10773-020-04604-z}
\BIBentrySTDinterwordspacing

\bibitem{bib_qba}
\BIBentryALTinterwordspacing
M.~Fitzi, N.~Gisin, and U.~Maurer, ``{Quantum Solution to the Byzantine Agreement Problem},'' \emph{Physical Review Letters}, vol.~87, no.~21, nov 2001. [Online]. Available: \url{https://doi.org/10.1103/physrevlett.87.217901}
\BIBentrySTDinterwordspacing

\bibitem{bib_qkd}
\BIBentryALTinterwordspacing
P.-Y. Kong, ``{A Review of Quantum Key Distribution Protocols in the Perspective of Smart Grid Communication Security},'' \emph{IEEE Systems Journal}, pp. 1--14, 2020. [Online]. Available: \url{https://doi.org/10.1109/JSYST.2020.3024956}
\BIBentrySTDinterwordspacing

\bibitem{Li2013}
\BIBentryALTinterwordspacing
J.~Li, L.~Li, H.~Jin, and R.~Li, ``{Security analysis of the “Ping–Pong” quantum communication protocol in the presence of collective-rotation noise},'' \emph{Physics Letters A}, vol. 377, no.~39, p. 2729–2734, nov 2013. [Online]. Available: \url{http://dx.doi.org/10.1016/j.physleta.2013.08.019}
\BIBentrySTDinterwordspacing

\bibitem{Liu2024}
\BIBentryALTinterwordspacing
A.~Liu, Q.~Zhang, S.~Xu, H.~Feng, X.-b. Chen, and W.~Liu, ``{QBIoT: A Quantum Blockchain Framework for IoT with an Improved Proof-of-Authority Consensus Algorithm and a Public-Key Quantum Signature},'' \emph{Computers, Materials \&; Continua}, vol.~80, no.~1, p. 1727–1751, 2024. [Online]. Available: \url{http://dx.doi.org/10.32604/cmc.2024.051233}
\BIBentrySTDinterwordspacing

\bibitem{Wazid2024}
\BIBentryALTinterwordspacing
M.~Wazid, A.~K. Das, and Y.~Park, ``{Generic Quantum Blockchain-Envisioned Security Framework for IoT Environment: Architecture, Security Benefits and Future Research},'' \emph{IEEE Open Journal of the Computer Society}, vol.~5, p. 248–267, 2024. [Online]. Available: \url{http://dx.doi.org/10.1109/OJCS.2024.3397307}
\BIBentrySTDinterwordspacing

\bibitem{Jain2023}
\BIBentryALTinterwordspacing
S.~Jain, R.~Aggarwal, and A.~B. Gandhi, ``{Quantum chain of things: “the quantum triad - integrating blockchain, IoT and quantum computing for a transcendent future”},'' \emph{IET Conference Proceedings}, vol. 2023, no.~22, p. 299–304, nov 2023. [Online]. Available: \url{http://dx.doi.org/10.1049/icp.2023.2892}
\BIBentrySTDinterwordspacing

\end{thebibliography}

\vfill


\end{document}